\documentclass[openacc]{rstransa}

\begin{document}

\title{Hilbert's Sixth Problem: \\ the endless road to rigour}

\author{
A. N. Gorban$^{1}$}

\address{$^{1}$Department of Mathematics,
University of Leicester, Leicester
LE1 7RH, UK}

\subject{Mathematics, Physics, Logic, Philosophy}


\corres{A. N. Gorban\\
\email{a.n.gorban@le.ac.uk}}

\begin{abstract}
Introduction, where the essence of the Sixth Problem is discussed and the content of this issue is introduced.
\end{abstract}


\begin{fmtext}

\section{The Sixth Problem}
In the year 1900 Hilbert presented his problems to the International Congress of Mathematicians (he presented 10 problems at the talk; the full list of 23 problems was published later). The list of 23 Hilbert's problems was very influential for 20th century mathematics. The Sixth Problem concerns the axiomatization of those parts of physics which are ready for a rigorous mathematical approach.
Hilbert's original formulation (in English trans\-la\-tion) was:

``6. Mathematical Treatment of the Axioms of Physics. The investigations on the foundations of geometry suggest the problem: To treat in the same manner, by means of axioms, those physical sciences in which already today mathematics plays an important part; in the first rank are the theory of probabilities and mechanics.'' This is definitely ``a programmatic call'' for the axiomatization of existent physical theories.

In a further explanation Hilbert proposed two specific problems: (i) axiomatic treatment of probability with limit theorems for the foundation of statistical physics, and (ii) the rigorous theory of limiting processes ``which lead from the atomistic view to the laws of motion of continua'':
"As to the axioms of the theory of probabilities, it seems to me desirable that their logical investigation should be accompanied by a rigorous and satisfactory development of the method of mean values in mathematical physics, and in particular in the kinetic theory of gases. ... Boltzmann's work on the principles of mechanics suggests the problem of developing mathematically the limiting processes, there merely indicated, which lead from the atomistic view to the laws of motion of continua."

The Sixths Problem has inspired several waves of research. Its mathematical content changes in time in a way that is very natural for a ``programmatic call'' \cite{1}.

\end{fmtext}
\maketitle

\begin{figure}
\centering
\includegraphics[width=\textwidth]{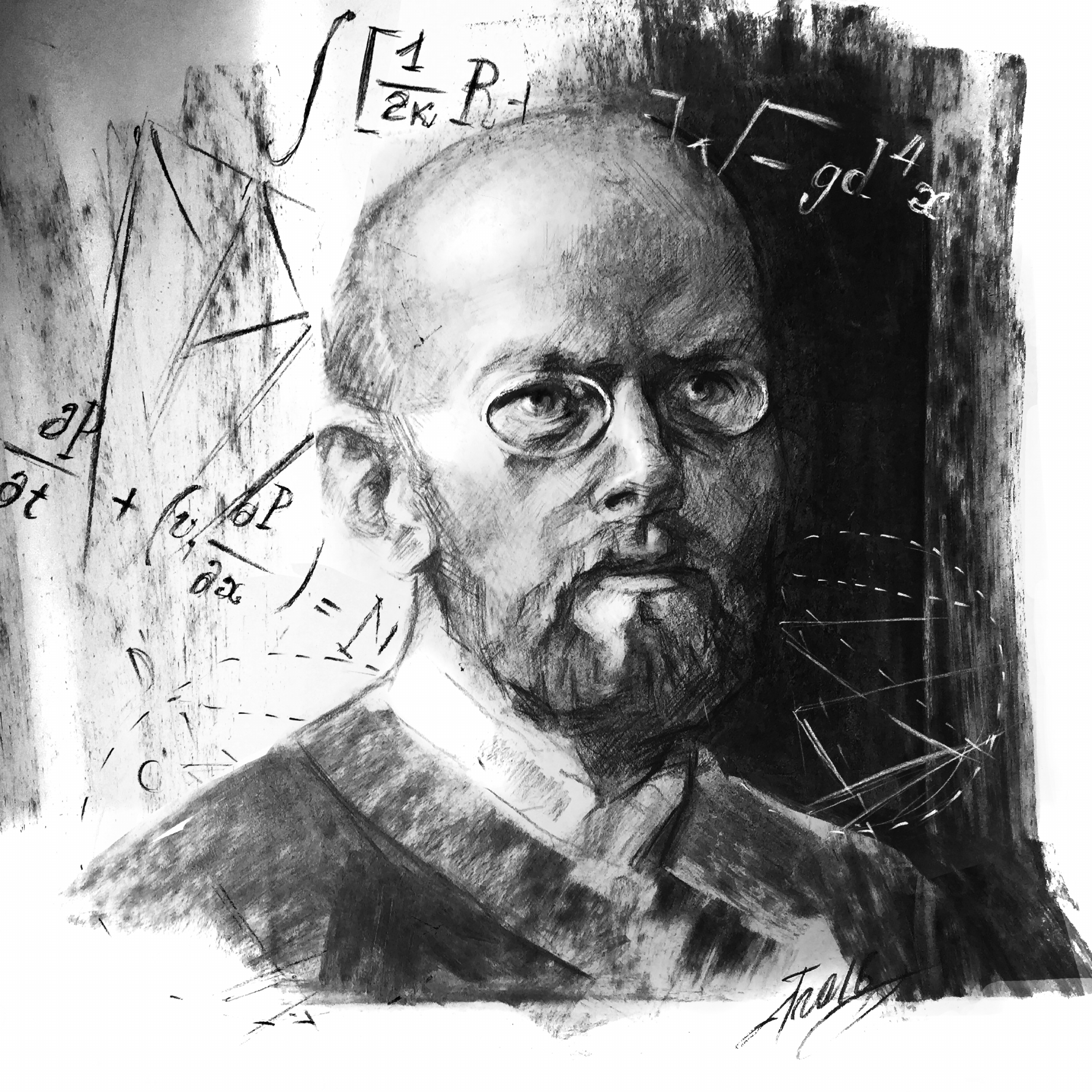}
\caption {David Hilbert, the great reformer of  science. The cover image of the issue. Courtesy the artist \href{http://annagorban.ru/}{Anna Gorban}, 2018}
\label{Hilbert}
\end{figure}

In the 1930s, the axiomatic foundation of probability seemed to be finalized on the basis of measure theory \cite{2}. Nevertheless, Kolmogorov and Solomonoff in the 1960s stimulated new interest in the foundation of probability (Algorithmic Probability) \cite{3}.

Hilbert, Chapman, and Enskog created  asymptotic expansions for the hydrodynamic limit of the Boltzmann equations \cite{4,5}. The higher terms of the Chapman-Enskog expansion are singular and truncation of this expansion does not have rigorous sense (Bobylev \cite{6}). Golse, Bardos, Levermore and Saint-Raymond proved rigorously the Euler limit of the Boltzmann equation in the scaling limit of very smooth flows \cite{7,8}, but recently Slemrod used the exact results of Karlin and Gorban \cite{9,10} and proposed a new  Korteweg asymptotic of the Boltzmann equation \cite{11}. These works attracted much attention and a special issue of BAMS was published with papers of L. Saint-Raymond \cite{12}, and A.N. Gorban and I. Karlin \cite{13}. There was a media reaction: QuantaMagazine published on July 21, 2015 a paper ``Famous Fluid Equations Are Incomplete: A 115-year effort to bridge the particle and fluid descriptions of nature has led mathematicians to an unexpected answer.''

It seems that Hilbert presumed the kinetic level of description (the ``Boltzmann level'') as an intermediate step between the microscopic mechanical description and the continuum mechanics. Nevertheless, this intermediate description may be omitted. Now, L. Saint-Raymond with co-authors is developing a new approach to the problem ``from the atomistic view to the laws of motion of continua'' without intermediate kinetic equations \cite{14}.

Quantum mechanics was invented after the Hilbert problems were stated. Almost simultaneously with the birth of quantum theory (in 1925), he devoted a seminar to the description of its mathematical structure. The notes of this seminar, collected by von Neumann and Nordheim (Hilbert's assistants at that time), were later published in a joint paper \cite{15}. In this paper the authors declare that the goal of an axiomatization of a physical discipline consists in: ``... formulating the physical requirements so clearly, that the mathematical model becomes uniquely determined by them ...''. The main idea is that a physical theory consists of three, sharply distinguishable parts: (i) physical axioms, (ii) analytic machinery (also called ``formalism''), and (iii) physical interpretation. The first attempt at formalization of quantum mechanics was performed by von Neumann \cite{16}.

The Kolmogorov formalization of the classical probabilistic model was published in a time (1933) when quantum mechanics was already making extensive use of a completely different probabilistic formalism. In the following 20 years each of the two disciplines was strongly concentrated on its own inner development and we had to wait until the second Berkeley Symposium on Probability and Statistics (1950) when Feynman for the first time explained to a large and qualified international audience of probabilists, the scientific challenge posed by the existence of two, apparently mutually incompatible, mathematical models of probability theory \cite{17}.

After that, the axiomatic approach to  quantum probability was developed by many researchers \cite{18} and there are many versions of its axiomatization \cite{19,20}. The fast development of this area of research has involved ideas from algebra, functional analysis, and fuzzy set theory. There has even developed a programming language based on non-commutative logic \cite{21}. Ideas and methods of quantum computing \cite{22,23} and quantum cryptography \cite{24} transform research in the foundation of quantum mechanics into an applied discipline with a perspective of applications in engineering. Many new mathematical structures and methods have been invented (see, for example, \cite{25,26}). Applications of quantum probability are now much wider than just the mechanics of subatomic particles \cite{27}.

Work on Hilbert's Sixth Problem involves many areas of mathematics: mathematical logic, algebra, functional analysis, differential equations, geometry, probability theory and random processes, theory of algorithms and computational complexity, and many others. It remains one of the most seminal area of interdisciplinary dialog in mathematics and mathematical physics.

 \section{This issue}

In the first paper \cite{CorryPTRS} L. Corry explains  the essence of the Sixth Problem as a programmatic call for the axiomatization of the physical sciences. Then two reviews follow. R. Hudson \cite{HudsonPTRS} gives a survey of the ``non-commutative'' aspects of quantum probability related to the Heisenberg commutation relation. L. Accardi \cite{AccardiPTRS} explains that ``One can say that, with the birth of quantum theory, Hilbert's Sixth Problem was split into three different questions:
\begin{enumerate}
  \item Axiomatize classical probalility.
  \item Axiomatize the new (quantum) probalility.
  \item Clarify the connections between the two.''
\end{enumerate}
He presents systematically the Kolmogorov compatibility conditions, the differences between Non-Kolmogorovian and Kolmogorovian models, and the information theoretic formulation of Heisenberg principle. Emergence of Hilbert spaces, the Schr{\"o}dinger equations, and gauge theories is demonstrated.

Lozada Aguilar, Khrennikov, and Oleschko in their `opinion piece' paper \cite{KhrennikovPTRS} move the discussion much further and present  a nonclassical application of nonclassical probability. They go from axiomatics of quantum probability and the theory of open quantum systems to modelling of geological uncertainty and management of intelligent hydrocarbon reservoirs.

F. Golse \cite{GolsePTRS} develops a bridge between the quantum and classical parts of Hilbert's Sixth Problem: he studies the semiclassical  mean-field limit for the quantum N-body problem and finds a convergence rate for the mean-field limit. This work is closely related to the analysis of ``the limiting processes, there merely indicated, which lead from the atomistic view to the laws of motion of continua'', requested by Hilbert.

S. Majid in his  opinion piece \cite{MajidPTRS} goes beyond the commonly accepted quantum theories. The mission of his approach is to be a step towards understanding ``how both General Relativity and Quantum Theory could emerge from some deeper theory of quantum
gravity''. He also reviews some previous works and proposes a simple toy 4-point model, which aims to give a hint about quantum gravity in space-time.

The series of works devoted to the ``satisfactory development of the method of mean values in mathematical physics, and in particular in the kinetic theory of gases'' is opened by the paper of A. Bobylev \cite{BobylevPTRS}. He considers the problem of higher equations of hydrodynamics. For this purpose, he proposes and analyses the method of successive changes of hydrodynamic variables and estimates  the accuracy of the Navier-Stokes and higher approximations.
The problem of the proper reduction from kinetics to fluid dynamics, stated by Hilbert in 1900, is not yet solved. New equations of continuum mechanics between Navier--Stokes and Boltzmann equations may be needed.  M. Slemrod \cite{SlemrodPTRS} states that the classical ``Boltzmann to Euler limit'' fails in the most interesting cases, and the equations should be changed. He uses as the prototype the exactly solved reduction problem for simplified kinetics \cite{9,13} and proposes the Korteweg fluid dynamics for the non-equilibrium flows.  L.G. Margolin \cite{MargolinPTRS} also aims to derive new fluid dynamic equations. He uses the local averaging in the coordinate space and derives the hydrodynamics models for the averaged moments. I. Karlin \cite{KarlinPTRS} demonstrates how to apply the regular principle  of dynamic invariance for regularisation of Grad's moment system. He finds and deciphers the detailed structure of the regularised 13 moment system.
R. MacKay \cite{MacKayPTRS} considers the reduction problem for Markov processes in more formal mathematical settings and proposes a new procedure for such reduction.

A.N. Gorban and I.Y. Tyukin \cite{GorbanPTRS} show how the programmatic call of Hilbert's Sixth problem influences statistical mechanics and machine learning. They  present new aspects of the concentration of measure phenomena, the stochastic separation theorems, and new  non-destructive procedures for correction of unavoidable errors for artificial intelligence systems.

The free style opinion discussion of a possible full mathematisation of  physical theories and a novel
algorithmic paradigm for physics is presented by G.M. D'Ariano \cite{D'ArianoPTRS}.

\section{The problem of reality: what we keep silence about}

{\em Semantics of physics} should be different both from the model-theoretical semantics of formal theories and from the various linguistic and epistemic semantics.  Development of semantic of physical reality is a non-trivial task: it should have sufficient level or rigour for use in theoretical physics and  sufficient level of generality for work with the wide range of physical theories, including yet unknown future physics.

Here we briefly  observe, from a bird's eye view, the problem of semantics associated with the Sixth Problem. For discussion of semantic we employ  the idea of possible worlds in its wide sense.
The {\em possible world} of a theory is a possible course of events, with full details, which is fully concordant with the theory.

Because of our education, our minds live in a  bizarre configuration of possible worlds of physical theories. Mechanics (with continuum mechanics), electromagnetism, and optics cover our everyday experience. If we meet anything different, we refer to other physical disciplines: statistical physics, quantum mechanics, and so on.

The idea that there exists a general theory, the truth, which rules the world, is a very powerful driving force for theoretical physics. Einstein, for most of his life, tried to approach such a theory. He definitely knew that the existence of this theory cannot be proved but it can just be postulated. In his famous discussion with R. Tagor \cite{GhoseEinsteinTagor2017}, Einstein formulated his point of view unambiguously:
\begin{itemize}
\item[] EINSTEIN: I cannot prove, but I believe in the Pythagorean argument, that the truth is independent of human beings. It is
the problem of the logic of continuity.
\item[] ..........................
\item[ ] TAGORE: In any case, if there be any truth absolutely unrelated to humanity, then for us it is absolutely non-existing.
\item[ ] EINSTEIN: Then I am more religious than you are!
\end{itemize}
It is absolutely necessary to mention here for clarity that Einstein often expressed scepticism with regard to the traditional creeds and God for him was similar to the God of Baruch Spinoza. (His sentence ``I believe in Spinoza's God, who reveals himself in the harmony of all that exists, ...'' became well-known \cite{CalapriceEinstein2010}). For Spinoza, God was {\em natura naturans} (nature doing what nature does) \cite{JaspersSpinoza}.

We can call the belief in the rational construction behind the Universe, the {\em ultimate rationalism}. The idea of ``ultimate rationalism which urges forward science and philosophy alike'' belongs to Whitehead \cite{Whitehead1929}.

I propose to take seriously the title of the Sixth Problem: {\em ``Mathematical Treatment of the Axioms of Physics''}. Of course, the explanation that follows the headline, reduces the sublime sound of the title. But we should take into account that Hilbert was  a highly qualified scientist and an experienced science writer. He understood the potential reaction of the readers of this title, no doubt. This could not be done unintentionally.

If we consider this problem seriously then we should be ready to assume that the real Universe is a possible world of a theory (yet unknown) and our goal is to reveal this theory. Einstein insisted that the proper theory should be categorical as much as it is possible: ideally, the structure described by the theory should be practically unique. (He discussed relations between the equations, which are fixed by the theory, and the initial conditions, which are arbitrary,  and proposed to decrease this freedom of choice  by additional theoretical principles like the Mach principle.)

Alas, it is quite possible that we aim to create the {\em ``theory of everything''} but instead of a beautiful theory always approach a system of {\em ``standard models''}, which do not pretend to be the truth and, possibly, have a hidden contradiction, either in the theory, or with the reality. It may happen that instead of the possible world of a nice theory, our world will be always an {\em ``impossible possible world''} with hidden contradictions \cite{Hintikka1979}.

The modal and epistemic logics consider the contradictions and impossible worlds as epistemic problems, whereas in `reality' there exists the truth and non-contradictory semantics. Here we allow another possibility: there is no non-contradictory theory of everything (even hidden), and the world has no rational reconstruction. We also cannot exclude the possibility that the rational truth is so well-hidden that we cannot reach it. The apparent result should be the same: standard models and hidden contradictions.

The web of standard models, which contradict each other  and have known mismatches with  reality, is the everyday practice of applied science. The ultimate rationalism is a driver that moves us to new, beautiful and simple theories. Nobody can state that it will release us from the web of standard models and from the mousetraps of impossible possible worlds, but the dream helps and gives energy.

Another problem with the semantics of physics is that there exists no atheoretical formalisation of a notion of physical {\em reality}. The possible worlds of theoretical physics consist of {\em things, their attributes and relations} prescribed by the theory. We have no semantics of physics without physics. The proposal of   D'Ariano to discuss ``physics without physics'' \cite{D'ArianoPTRS} is not just an elegant expression. We need the universal semantics. The  attempts from the beginning of the Twentieth Century to exclude theoretical terms and work with the language that reflects the everyday human experience are a bit na\"{\i}ve. Everybody who has experience in robotic vision knows that it is a highly non-trivial task to teach a robot the everyday experience of scene analysis and pattern recognition. Everyday natural language includes constructs, which are also far from direct perception.

We can try to outline the possible ``semantic of physical reality'' that complements semantics of physical theories, is simple, abstract, and free from  theoretical allusions. Let us just follow the machine learning abstractions. There are two alphabets: Perceptions $\mathcal{P}$ and Actions $\mathcal{A}$. The  ``person'' selects and sends ``outwards'' the symbols from $\mathcal{A}$ (actions) and receives symbols from $\mathcal{P}$ (the answers to actions,  or just signals from ``outsides''). There is the ``freedom of will'': selection of a symbol from  $\mathcal{A}$ has no restrictions ($\mathcal{A}$  models   commands to actuators, the real actions and their results return to the person with the $\mathcal{P}$-feedback).

The outer world can be modelled in any theory, can act as a classical or quantum device, etc. The person can learn, explore the world,   create theories and identify  models. At the end, the sets of $\mathcal{A}-\mathcal{P}$ sequences will correspond to sets of structures and models.

We should not expect that  such learning is driven by a utility optimisation. Quite the opposite, this learning can be goal-free. Gromov \cite{Gromov2011} identified {\em structure = interesting structure} and declared that {\em the goal free structure learning is a structurally interesting process.}

This is just a brief outline of the semantic problem related to the Sixth Problem:
\begin{itemize}
\item The problem of the ultimate rationalism versus the web of standard models and the possible impossible worlds;
\item The problem of a semantics of reality, which does not depend on a theory and is not related to  human physiology  or natural language.  The approach  considered is borrowed from machine learning;
\item The problem of structure discovery, model identification and goal-free learning.
\end{itemize}

\section{Conclusion}

Hilbert started his ``road to formal rigour'' from ``The Foundations of Geometry'' \cite{Hilbert1902a}. He expected, that in an axiomatic form, the theory would be developed independently of any need for intuition: ``Professor Hilbert has, so to speak, sought to put the axioms into such a form that they might be applied by a person who would not understand their meaning because he had never seen either point or straight line or plane'' \cite{Poincare1903}.

In 1900, Hilbert's axiomatizing project was well accepted by the community of mathematicians. Some minor comments about incompleteness of his ``Foundations'' required just incremental work. G{\"o}del  was not yet born. Hilbert's Sixth Problem  was the declaration of the expansion of the axiomatic method outside the existing mathematical disciplines, in physics and beyond.

There are three general lessons from the history of the Sixth Problem:
\begin{itemize}
\item Attempts at axiomatizing are useful even beyond mathematics: they help to clarify the basic assumptions and keep the mind in order. This is good and encouraging.
\item This order may be much more complicated and counter-intuitive than  the preceding chaos. This is not so encouraging, but not too bad: axiomatizing is a   powerful tool, and its use requires accuracy.
\item The road to rigour is infinite. This is neither encouraging nor discouraging: this is reality.
\end{itemize}

Discussions of Hilbert's Sixth Problem and preparation of this issue gave us one more lesson: The road to rigour is infinite {\em and interesting}. Some comments by the Workshop participants are collected after the bibliography.

\appendix

\section{Opinions of workshop participants}

\noindent ``This workshop was rather unique in its kind as it brought together mathematicians from very different horizons (from probability theory to statistical physics, partial differential equations, logic and complexity), whose common interest lies in Hilbert's sixth problem. The statement of Hilbert's sixth problem is indeed so broad that it can be studied from a multitude of angles, and it was really extremely interesting for me to hear some of the best specialists from those various fields explain their viewpoint of the problem, and the progress that has been made recently. I think all the participants share this impression on having gone on a wonderful tour of the sixth problem for three days, and have come home with new ideas, and lots of new questions.
\vspace{-4mm}
\begin{flushright}
Isabelle Gallagher

Universit{\'e} Paris-Diderot (Paris 7),

UFR de Math{\'e}matiques,

Paris, France.
\end{flushright}

\noindent ``As a historian of mathematics who has devoted many years of his academic life to research the origins, the development and the overall impact of Hilbert's sixth problem, this conference has been for me a unique kind of scholarly experience. Historians of mathematics are used to work in intellectual isolation. Typically, our work is too technical and daunting for historians of science in general. At the same our work is too ``historical'' for what should be one of our main natural audiences, namely, that of the mathematicians (and in the case of Hilbert's sixth problem, also the physicists). I have had previous opportunities to speak to mathematical audiences about my work on the problem, but with special attention to the encounter between Hilbert and Einstein around the final formulation of the field equations of general relativity. Quite unsurprisingly, this has proved to be an appealing enough topic for such audiences. But this conference was fully devoted to the much broader topics associated with the Hilbert problem, and to a truly wide range of topics, and it was very refreshing to realize the extent to which the ideas put forward by Hilbert more than 115 years proved to be much more relevant and of long-term impact than Hilbert himself could have ever conceived when formulating his programmatic call for the axiomatization of physical theories in 1900. The talks were not only informative but also inspiring, and for me they were an eye opener in many directions. I was also glad to come out with the feeling that my own historical talk arose significant interest among the attendees.
\vspace{-4mm}
 \begin{flushright}

Leo Corry,

The Lester and Sally Entin Faculty of Humanities,

Tel Aviv University

Israel
\end{flushright}

\noindent ``Hilbert was one of the most successful leaders to set foundations of mathematics by axiomatic methods.
  Hilbert's sixth problem called for extending this method out of mathematics to physical science.  Mathematicians who made significant contributions to this problem in the last decades world wide gathered and discussed the significant roles of the problem playing mainly in hydromechanics and quantum physics.
  The workshop revealed the unexhaustive value of the problem, which has enriched both physical science and mathematics for more than 100 years and will lead to unexpected mutual interactions between those intellectual arts.
\vspace{-4mm}
\begin{flushright}
Masanao Ozawa

Graduate School of Information Science

Nagoya University

Japan
\end{flushright}

\noindent ``The Workshop on ``Hilbert's Sixth Problem'' was an inspriring experience that shed a new light on the problem itself and on the approaches of its solution. The carefully selected lectures from both the classical and quantum problems of mathematical physics have clearly showed that Hilbert's Sixth Problem has not been completely solved but presents a lot of interesting research questions to be answered even today not only in applied mathematics and physics but also in engineering.   Thanks to the excellent organization and to the friendly and relaxed atmosphere there were plenty of opportunities for discussions and hopefully cross-fertilization of the ideas and solution methods in the interdisciplinary area between applied mathematics, physics and a bit of engineering.
\vspace{-4mm}
\begin{flushright}
Katalin Hangos

Process Control Research Group

Computer and Automation Research Institute

Hungarian Academy of Sciences

Budapest, Hungary
\end{flushright}

\noindent ``In 1900, the German mathematician David Hilbert laid out a program (known as Hilbert's 6th problem) calling for no less than the ``axiomatization of physics''. As an example, Hilbert proposed to derive the classical equations of fluid mechanics from the atomistic description of gases --- at a time when even the existence of atoms was a matter of heated scientific debate. Since then, new paradigms have emerged in physics: relativity, and quantum mechanics have altered our understanding of the world. Today, the quest for a mathematical coherence in modern physics is even more formidable a challenge than in the days of Hilbert. The workshop organized by Prof. A. Gorban and his colleagues at the University of Leicester has been a unique opportunity to assemble a panel of mathematical physicists from various countries (including for instance China, France, Italy, Russia, Sweden, Switzerland, in addition to the UK), with a great variety of viewpoints on Hilbert's 6th problem.
\vspace{-4mm}
 \begin{flushright}
Fran\c{c}ois Golse

D{\'e}partement de Math{\'e}matiques

{\'E}cole Polytechnique,

Paris, France
\end{flushright}

\noindent ``The Workshop on ``Hilbert's Sixth Problem'' organized by Alexander Gorban in Leicester will have a great impact on the community of physicists and mathematicians. For the first time a workshop has been devoted to a problem of the outmost relevance at the methodological level for the advancement of theoretical physics, in a time of great evolution of theoretical physics. Indeed, the solution of the VI Hilbert problem, namely the axiomatization of physics, will play a pivotal role in finding the correct answer to the major problems in contemporary physics, including the solution of the logical clash between General Relativity and Quantum Theory, and that of the emergence of Hydrodynamics from Statistical Mechanics. We have attended very interesting talks during this conference, and we have come back home with a todo list of relevant issues to address and a set of methodological recipes for improve the quality of theoretical research. David Hilbert has changed the history of physics with his research and his teaching in G{\"o}ttingen during the quantum and relativity revolution, and he is still teaching to us nowadays the most important lesson: that the long-lasting relevant science must be built on solid foundations.
\vspace{-4mm}
 \begin{flushright}

Giacomo Mauro D'Ariano

Istituto Nazionale di Fisica della Materia,

Unita' di Pavia,

Pavia, Italy
\end{flushright}

\noindent ``In 1900 David Hilbert provided the mathematicians of his day with a list of problems for the twentieth century and in particular his sixth problem which called for the axiomatization of physics in the same spirit as we would treat geometry. The problem in its various interpretations has provided a fertile ground for mathematicians for over one hundred years with a rather impressive list of results and even unintended consequences. This conference (perhaps the first of its kind) expanded on the physics known to Hilbert in 1900 (statistical and continuum mechanics) and included the issues of the very small (quantum mechanics), very large (general relativity), and basic fundamental issues of probability which is the key component of both the statistical and quantum view of nature. At first thought it seems like a workshop with this broad view would have the difficulty of too large a perspective but the just the reverse was true. The participants excelled at thinking about the big picture and their presentations and conversations went far beyond any possible narrowness and self-interestedness. It was a remarkable event.
\vspace{-4mm}
\begin{flushright}
Marshall Slemrod

Dept of Mathematics,

University of Wisconsin,

Madison, WI USA
\end{flushright}

\vskip6pt

\enlargethispage{20pt}

\competing{The author declares that he has no competing interests.}

\funding{The Workshop  ``Hilbert's Sixth Problem'' (University of Leicester, May 02-04, 2016) and  preparation of this issue were supported by  EPSRC grant EP/N022653/1, LMS conference grant No: 11503, and IMA Small Grant  SGS47\_15}.

\ack{The idea of this issue was proposed during the LMS-EPSRC-IMA supported workshop (conference) ``Hilbert's Sixth Problem'' (Leicester, May 2016) organised by Alexander Gorban. The Scientific Committee of the Workshop was: Luigi Accardi, Alexander Bobylev, Pierre Degond, and Marshall Slemrod. Most of the papers in the issue grow from the talks at the Workshop. We are very grateful to all the participants for the inspiring talks and seminal discussion!}


\end{document}